\documentclass[11pt,preprint]{aastex}
\usepackage{graphics}
\usepackage{amssymb}
\usepackage{amsmath}
\usepackage[dvips]{color}
\newcommand{\kms}{{km~s$^{-1}$}}
\newcommand{\fex}{[Fe {\sc X}]$\lambda$6376}
\newcommand{\fexi}{[Fe {\sc XI}]$\lambda$7894}
\newcommand{\fexiv}{[Fe {\sc XIV}]$\lambda$5304}
\newcommand{\fevii}{[Fe {\sc VII}]}
\newcommand{\arxiv}{[Ar {\sc XIV}]$\lambda$4414}
\newcommand{\sxii}{[S {\sc XII}]$\lambda$7612}
\newcommand{\oi}{[O {\sc I}]$\lambda$6300}
\newcommand{\oii}{[O {\sc II}]$\lambda$3727}

\newcommand{\nii}{[N {\sc II}]$\lambda$6583}

\newcommand{\sii}{[S {\sc II}]$\lambda$6716}
\newcommand{\siis}{[S {\sc II}]$\lambda$6731}
\newcommand{\oiii}{[O {\sc III}]$\lambda$5007}

\newcommand{\neiii}{[Ne {\sc III}]$\lambda$3896}
\newcommand{\hb}{H$\beta$}
\newcommand{\heii}{He {\sc II}$\lambda$4686}

\newcommand{\hii}{H {\sc II}}
\newcommand{\ha}{H$\alpha$}

\shorttitle{Extreme Coronal Line Emitters}
\shortauthors{Wang et al. 2011}
\received{2011 October 8}

\begin{document}

\title{Extreme Coronal Line Emitters: Tidal Disruption of Stars by 
       Massive Black Holes in Galactic Nuclei?}
\author{Ting-Gui Wang\altaffilmark{1,2}, Hong-Yan Zhou\altaffilmark{1,2,3}, 
        S. Komossa\altaffilmark{4,5,6,7}, Hui-Yuan Wang\altaffilmark{1,2}, 
        Weimin Yuan\altaffilmark{7}, and Chenwei Yang\altaffilmark{1,2}}
\altaffiltext{1}{Key Laboratory for Research in
Galaxies and Cosmology, The University of Sciences and Technology of
China (USTC), Chinese Academy of Sciences, Hefei, Anhui 230026, China;
~twang@ustc.edu.cn}
\altaffiltext{2}{Center for Astrophysics, USTC, Hefei, Anhui 230026, China}
\altaffiltext{3}{Polar Research Institute of China,451 Jinqiao Road, Pudong, 
                 Shanghai 200136, China}
\altaffiltext{4}{Technische Universit\"at M\"unchen, Fakult\"at f\"ur Physik, 
Lehrstuhl f\"ur Physik I, James Franck Strasse 1/I, 85748 Garching, Germany}
\altaffiltext{5}{Excellence Cluster Universe, TUM, Boltzmannstrasse 2, 85748 Garching, Germany}
\altaffiltext{6}{Max Planck Institut f¨ur Plasmaphysik, Boltzmannstrasse 2, 85748 Garching, Germany}
\altaffiltext{7}{National Astronomical Observatories, Chinese Academy of Sciences,
                20A Datun Road, Chaoyang District, Beijing 100012, China}

\begin{abstract}
Tidal disruption of stars by supermassive black holes at the centers of galaxies is
expected to produce unique emission line signatures, which have not yet been explored
adequately. Here we report the discovery of extremely strong coronal lines from [Fe {\sc X}]
up to [Fe {\sc XIV}] in a sample of seven galaxies (including two recently reported cases),
which we interpret as such signatures. This is the first systematic search for objects of
this kind, by making use of the immense database of the Sloan Digital Sky Survey.
The galaxies, which are non-active as evidenced by the narrow-line ratios, show broad
emission lines of complex profiles in more than half of the sample. Both the high
ionization coronal lines and the broad lines turn out to be fading on timescales of years
in objects observed with spectroscopic follow-ups, suggesting their transient nature.
Variations of inferred non-stellar continua, which have absolute magnitudes of at least
$-$16 to $-$18 mag in the $g$ band, are also detected in more than half of the sample. The
coronal line emitters reside in sub-L$_*$ disk galaxies ($-21.5 < M_i < -18.5$) with small
stellar velocity dispersions. The sample seems to form two distinct types based on the
presence or absence of the [Fe {\sc VII}] lines, with the latter having relatively low 
luminosities
of [O {\sc III}], [Fe {\sc XI}], and the host galaxies. These characteristics can most naturally be
understood in the context of transient accretion onto intermediate mass black holes at galactic 
centers following tidal disruption of stars in a gas-rich environment. We estimate the incidence
of such events to be around $10^{-5}$ yr$^{-1}$ for a galaxy with $-21.5 < M_i < -18.5$.

\end{abstract}

\section{Introduction}

It is generally accepted that nearly all massive galaxies host supermassive black holes at their
centers, and that the black hole masses are well correlated with the stellar masses or velocity
dispersions of the galactic bulges (Magorrian et al. 1998; Ferrares \& Merritt 2000; Gebhardt et
al. 2000). In the low mass regime, detections of active galactic nuclei (AGNs) in a small number
of dwarf or bulgeless galaxies suggest that black holes in the mass range of a few times 
$10^4-10^6$
$M_\sun$ do exist in such galaxies (Greene \& Ho 2004; Dong et al. 2007; Greene \& Ho 2007; Dong et al 2012). 
However, it is not clear how common they are among quiescent galaxies, as they are difficult to be explored with
the stellar and gas kinematics currently used, given the small spheres of gravitational influence.
Tidal disruption of stars by massive black holes will provide a signpost for the presence of such
black holes in quiescent galaxies. Theoretically the event rate is estimated to be $10^{-4}-10^{-5}$ 
galaxy$^{-1}$ yr$^{-1}$ (Rees 1988; Magorrian \& Tremaine 1999), and might be highest in nucleated dwarf
galaxies (Wang \& Merritt 2004). Thus the detection of a significant number of such events can be
used to estimate the incidence of dormant black holes in small galaxies. Black hole tidal disruption
events (TDEs) have other astrophysical consequences as well. They may contribute significantly to the
present-day black hole mass growth in galaxies fainter than 10$^9$ L$_\sun$ (Magorrian \& Tremaine 1999;
Milosavljev\'ic et al. 2006; Brockamp et al. 2011). The ejected unbound
debris may have significant impact on the energetics balance in the nuclear environment of small
galaxies (Rees 1988). Follow-up observations of the highly variable, flaring continuum and line
emission may provide a way of exploring the gas and dust distribution of galactic nuclei (Komossa
et al. 2008), with a technique similar to reverberation mapping used in the study of the broad
emission line region of AGNs (Peterson 2007); they may also provide a way of measuring the spin
and mass of the black holes (Beloborodov et al. 1992; Gezari et al. 2009).

Several observational signatures were predicted by theories. A star is tidally disrupted as it
plunges into the tidal radius of a massive black hole, $r_p < R_T \simeq R_*(M_{\rm BH}/M_*)^{1/3}$ 
(Hills 1975).
Less than half of the debris falls back and forms an accretion torus around the black hole, giving
rise to a flare of electromagnetic radiation. The flare will last for a few months to a year with a
peak at the Eddington luminosity for a black hole mass less than 10$^7$ $M_\sun$ (e.g., Rees 1988). For
a black hole of mass in the range of 10$^7$-10$^8$ $M_\sun$ and a solar-type star, the peak luminosity will
be sub-Eddington. The optically thick radiation of the accretion disk is peaked in the extreme$-$UV
or soft X-rays. The exact peak frequency depends on whether the fallback material forms a thick
torus at the pericenter radius, or forms a thin disk extending to the least stable orbit, or launches
an optically thick wind around the disk.

The bound and unbound debris, ionized by radiation from the accretion torus, is predicted
to emit broad emission lines (e.g., Ulmer 1999; Eracleous et al. 1995; Strubbe \& Quataert 2009;
Clausen \& Eracleous 2011). Numerical simulations showed that the unbound debris is kinematically
complex, producing complicated and variable line profiles (Bogdanov\'ic et al. 2004). UV and X-ray
photons illuminate interstellar gas, resulting in narrow emission lines, which last even longer (e.g.,
Ulmer 1999). Due to the hard ionizing continuum, the emission line spectrum is characterized 
by strong high-ionization lines, such as \heii. 

So far, about a dozen candidates of TDEs were identified from X-ray,
UV and optical surveys (e.g., Komossa \& Bade 1999; Komossa et al. 2004; Gezari et al. 2009;
van Velzen et al. 2011; Cenko et al. 2012). Follow-up observations confirmed that the X-ray or
UV luminosity declines with time as $t^{-5/3}$ (e.g., Komossa \& Bade 1999; Halpern et al. 2004) 
as predicted by theoretical models. The peak luminosities are around $10^{42}-10^{48}$ erg s$^{-1}$ 
in soft X-rays and $10^{44}-10^{46}$ erg s$^{-1}$ in the ultraviolet. These observations are consistent 
with the scenario of the tidal disruption of a solar-type star by a supermassive black hole of a 
few $10^6 - 10^8$ M$_\sun$. However, no high-ionization emission line signatures have been 
reported from these X-ray and UV-discovered events. \footnote{Note, however, the cases 
of two {\em active} galaxies which showed highly variable high-ionization emission lines (Peterson \& Ferland 1986; Brandt et al. 1995; Grupe et al. 1995). The cause for line variability remained unknown, but a TDE was mentioned as one possibility.}  
Yet the emission lines 
would provide further diagnostics for TDEs, and may even provide a new means of 
detecting them.

Of particular interest, Komossa et al. (2008) discovered an extreme coronal line emitter (ECLE,
hereafter), SDSS J0952+2143, that showed strong \fex\ comparable to \oiii, from
the Sloan Digital Sky Survey (SDSS). Follow-up observations (with BAO 2.16 m and NTT) showed
that both the high-ionization coronal lines and broad emission lines were strongly fading (Komossa
et al. 2009). As soft X-ray photons are needed to produce the coronal lines (hereafter CLs), these
authors proposed that the CLs are echoes of a recent stellar tidal disruption flare, and the broad
lines are emitted from the unbound debris, photoionized by the radiation from the accretion disk
(Komossa et al. 2008).

Very recently, another ECLEs, SDSS J0748+4712, was discovered by Wang et al. (2011),
which shows strong high-ionization CLs, such as \fex, \fexi, \fexiv, \arxiv, and 
\sxii, but no \fevii. Interestingly it also shows, in addition to a blue
continuum, very broad bumps, which can be interpreted as the blueshifted He {\sc II} and Balmer lines
from a super-Eddington radiation driven wind (Strubbe \& Quataert 2009). By comparison of the
SDSS photometric and spectroscopic data, the authors could constrain that the SDSS spectrum
was taken at most 4 months after the peak of the flare. The CLs and broad features became
very weak or absent in spectra taken 4-5 years later, while the [O {\sc III}] line luminosity increased by
a factor of 10. Despite the strong CLs, the traditional line ratio diagnostics (the BPT diagram;
Baldwin et al. 1981) indicates that it is not an AGN. Although in that case the
possibility of observing a supernova (SN) cannot be completely ruled out, the scenario faces some
fatal difficulties: the CL luminosities are 1$-$2 orders of magnitude higher than those in SNe with
the most luminous CLs; there is a lack of low-ionization lines or He {\sc I} with similar line profiles to
CLs as usually seen in SNe with CLs; the broad bumps cannot be self-consistently explained. Thus
TDE is left as the most promising scenario (Wang et al. 2011).

In this paper, we present a comprehensive study of a small, uniform sample of ECLEs culled
from the spectroscopic data set of SDSS Data Release 7 (DR7; Abazajian et al. 2009). Our goal
is to explore their nature and to estimate the rate of such events. The two previously reported
events, SDSS J0952+2143 and SDSS J0748+4712, are also included. The sample selection and data reduction are described in Section 2. The properties of these ECLEs are presented in Section 3. We discuss the
nature of these objects in Section 4 and summarize our results in Section 5. Throughout the paper we adopt a
$\Lambda$ dominated cosmology with $H_0 = 72$ km s$^{-1}$ Mpc$^{-1}$, ­$\Omega_\Lambda= 0.7$, 
and ­$\Omega_M = 0.3$.

\section{Sample and Data Reduction}

We carried out a systematic search for ECLEs in the spectroscopic samples of galaxies and
low-redshift quasars from the SDSS Data Release 7. We restrict redshifts $z<0.38$ to ensure
that the main CLs, such as \fex, \fexiv\ and \fevii$\lambda$6088, and H$\alpha$ fall within the
spectral coverage. The continuum fitting is done separately for galaxies and quasars. For galaxies,
the continuum is fitted with six IC templates derived from the stellar populations with Ensemble
Learning Independent Component Analysis (Lu et al. 2006). For quasars and AGNs, a power-law
and Fe {\sc II} templates are added to model the AGN continuum and the Fe {\sc II} pseudo-continuum. We
adopt the I Zw 1 templates (V\'eron-Cetty et al. 2004) for the broad and narrow Fe {\sc II}
lines, which are modeled with a single Gaussian of the same width and same redshift for narrow
or broad components. This should be sufficient for narrow emission line measurements used in the
sample selection stage. In the continuum fit, prominent emission lines are masked, as well as bad
pixels.

After subtracting the continuum, we fit the emission line spectrum with a number of Gaussians:
one or two Gaussians for each CL, \oiii\ and \neiii, which often show an additional
broad wing; two or three Gaussians for narrow and broad H$\alpha$ and H$\beta$; one Gaussian for other
low-ionization or weak narrow lines, such as \nii, \oii, \oi, \sii, and \oiii. In the case of 
a multi-Gaussian fit, the addition of one more Gaussian is justified based
on an $F$-test. In the fit the line widths and centers of \oiii, \nii, \sii\, and
\oi\ doublets are tied, and the flux ratios of \oiii, \nii\, and \oi\
doublets are set to the theoretical values.

ECLEs are then selected through the following criteria: (1) at least one CL is detected at the
5$\sigma$ level; (2) for at least one CL the strength is more than 20\% of the \oiii\ strength, 
or it is not spectroscopically classified as AGN according to narrow line diagnostics. Note that 
the strength of the CLs is typically only a few percent of \oiii\ in Seyfert galaxies (e.g., 
Nagao et al. 2000). The selected spectra were then inspected manually to reject false detections. 
The final sample includes seven galaxies. Figure 1 shows the SDSS spectra of the objects. 

The spectra are then re-fitted with refined models. The continuum is now fitted with a combination of starlight 
and a power-law representing the contribution of a possible non-thermal component. The starlight is modeled as a linear combination of IC templates, which are broadened
to match the absorption line widths. The reddening of the starlight component is treated as a
uniform dust screen. A broadened Fe {\sc II} template is added in the spectrum of SDSS J1055+5637,
in which an excess at the optical Fe {\sc II} wavelengths is clearly visible. With these fits we also
obtain other parameters, such as the stellar velocity dispersions, the reddening of the stellar and
the Fe {\sc II} continua. Note that the power-law is not an independent component from the ICs, and
therefore the decomposition of the power-law and the starlight is not reliable. In the case of 
SDSS J0748+4712, the broad bumps are modeled as Gaussians (Wang et al. 2011). After the continuum
is subtracted, the emission line spectrum is re-fitted. While narrow lines are fitted with the same
models as before, broad lines are fitted with as many as three Gaussians. The H$\alpha$ and H$\beta$ 
lines are simultaneously fitted with models of the same profiles. An example of an emission line fit 
is shown in Figure \ref{fig2}. The basic data of the ECLE sample are tabulated in Table 1, and the emission line fluxes are listed in Table 2 and 3.

\section{Results}

\subsection{Continuum Variability}

We examine the possible continuum variability of the ECLEs by comparing their fiber magnitudes 
obtained in the SDSS imaging survey and the spectral magnitudes obtained in the spectroscopic 
survey. The two surveys were four months to two years apart. The fiber magnitudes are measured
within an aperture of the size of the SDSS fibers from the SDSS images assuming a typical seeing
of 2\farcs0. The spectral magnitudes are synthesized from the SDSS spectra, which are 
spectro-photometrically calibrated using standard stars in the three SDSS bands $g$, $r$, and $i$. 
This calibration can lead to a difference between fiber and spectral magnitude for extended sources 
even if there is no variability. This is because of different seeing during the spectroscopic 
observations, and object to object differences in radial profiles, and so there are both systematic 
offsets and random scatter, which vary from plate to plate. We estimate the mean offset and random 
scatter from the distributions of their offsets in the $g$, $r$, and $i$ bands using galaxies in 
the plates of the ECLEs (Figure \ref{fig3}), which can be well  fitted with one Gaussian (outliers 
are likely those containing AGNs). Based on the results, the probability and
amount of potential variations are estimated for the ECLEs, as listed in Table \ref{tab1}.

Four out of seven objects show significant ($\geq 3\sigma$) variations with amplitudes ranging from 0.1 to
0.3 mag in the $g$ band  and smaller at longer wavelengths) between the SDSS spectroscopic and
photometric observations. Assuming that the light at the lower state is solely from stars, we derive
the minimum brightness of a variable non-stellar component at the higher state, which ranges from
$-$16.0 to $-$18.0 mag in absolute magnitude in the $g$ band (Table \ref{tab1}). We note that two ECLEs without
\fevii\ (SDSS J0748+4712 and SDSS J1350+2916 in Table \ref{tab1}) brightened between the two observations while two with 
\fevii\ (SDSS J0952+2143 and SDSS J1055+5637) show the opposite trend.
In the former case, the interval between the two observations can set an upper limit on the ages of
the ECLEs during the spectroscopic observation; in the latter case, a lower limit can be obtained.
Consequently, the continuum flares occurred within 120 and 701 days prior to the spectroscopic
observation for SDSS J0748+4712 and SDSS J1350+2916, respectively, and happened more than 140 and 375
days prior to the spectroscopic observation for SDSS J1055+5637 and SDSS J0952+2143, respectively.

We have also checked archival UV data for potential variations. SDSS J0748+4712 was detected
by {\it Galaxy Evolution Explorer} in both the near- and far-UV bands on 
2004 March 10, 20 days after the SDSS spectroscopic
observation, and was re-observed in near-UV later on 2010 January 9. The NUV flux dropped by nearly
a factor of two at the later epoch. This sets a lower limit to the non-stellar UV luminosity as
2.6$\times$10$^{42}$ erg cm$^{-2}$ s$^{-1}$ on 2004 March 10, assuming that the flux at the later 
epoch is mostly from stellar light.

\subsection{Narrow Emission Lines}

\begin{figure}
\plotone{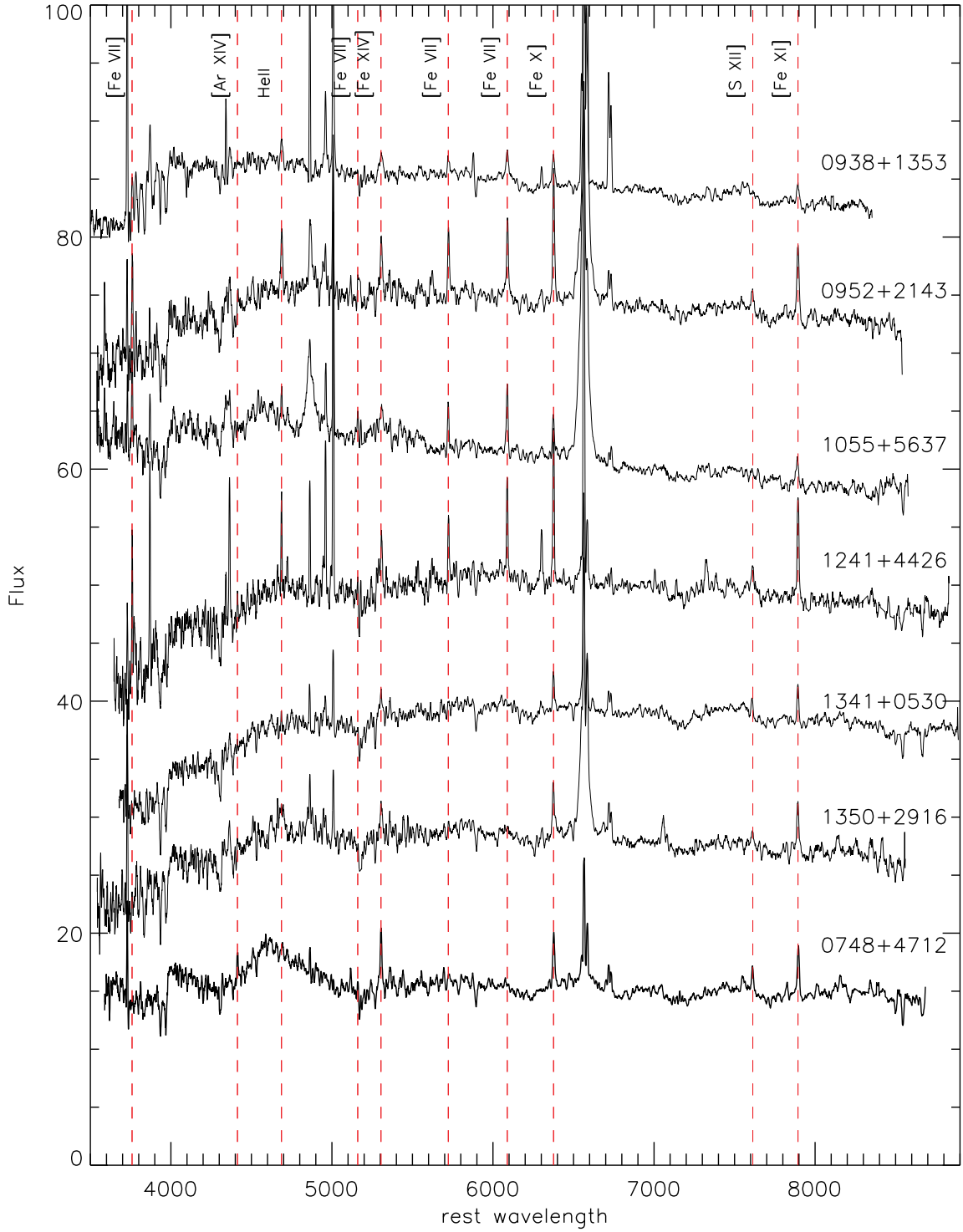}
\caption{SDSS spectra of the ECLEs presented in this paper. Coronal lines are marked in
red. The spectra have been shifted in vertical direction for clarity.}
\label{fig1}
\end{figure}

\begin{figure}
\plotone{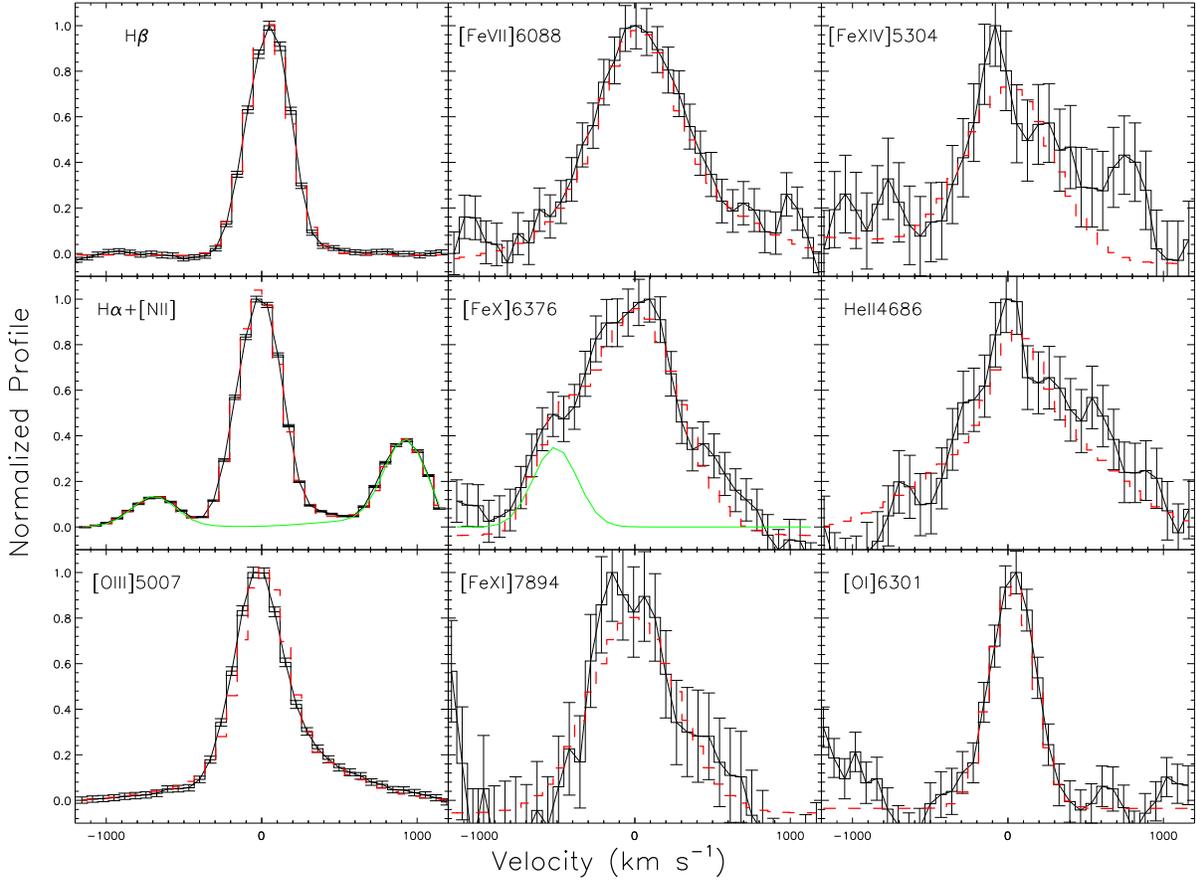}
\caption{Example of emission-line fits for strong coronal lines, 
[O {\sc III}]$\lambda$5007, He {\sc II}, \oi\, and Balmer lines of SDSS J0938+1353. 
The peak of the 
emission line has been normalized. The black solid line in each panel 
represents the observed profile, the red dashed line the fit, 
and the green line the blended other lines. }
\label{fig2}
\end{figure}

\begin{figure}
\plotone{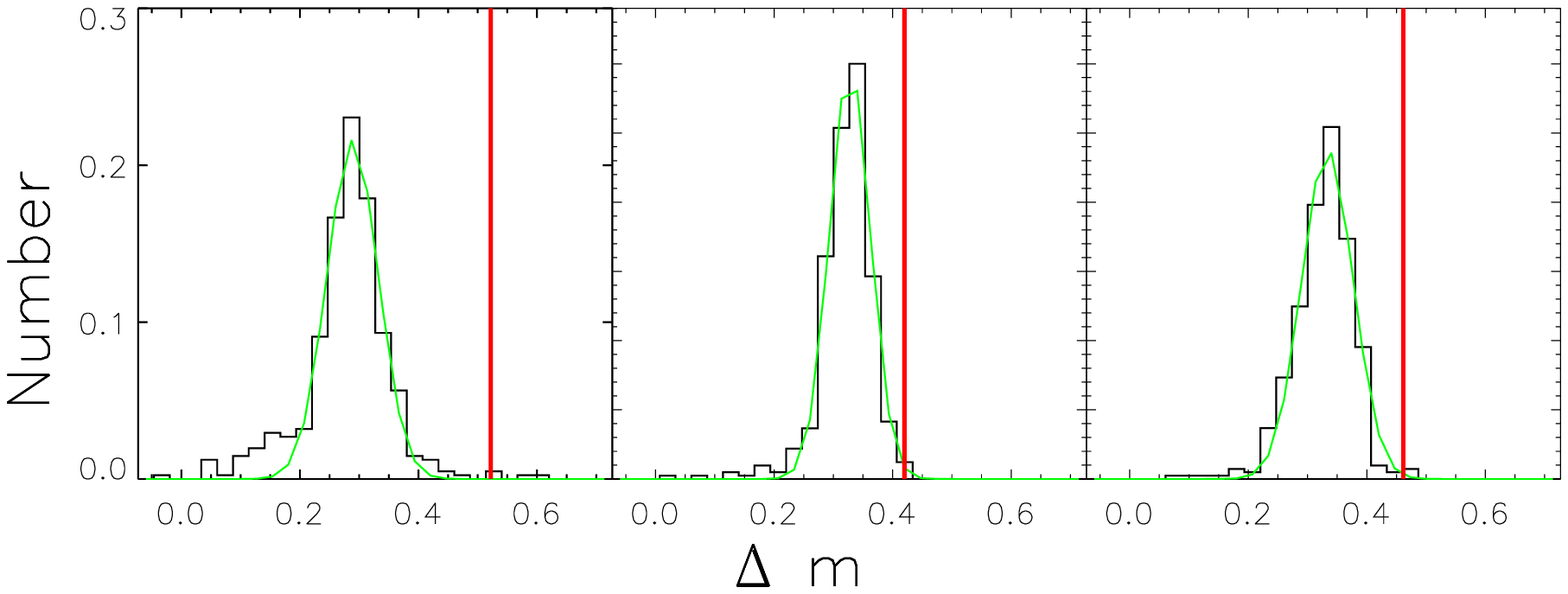}
\plotone{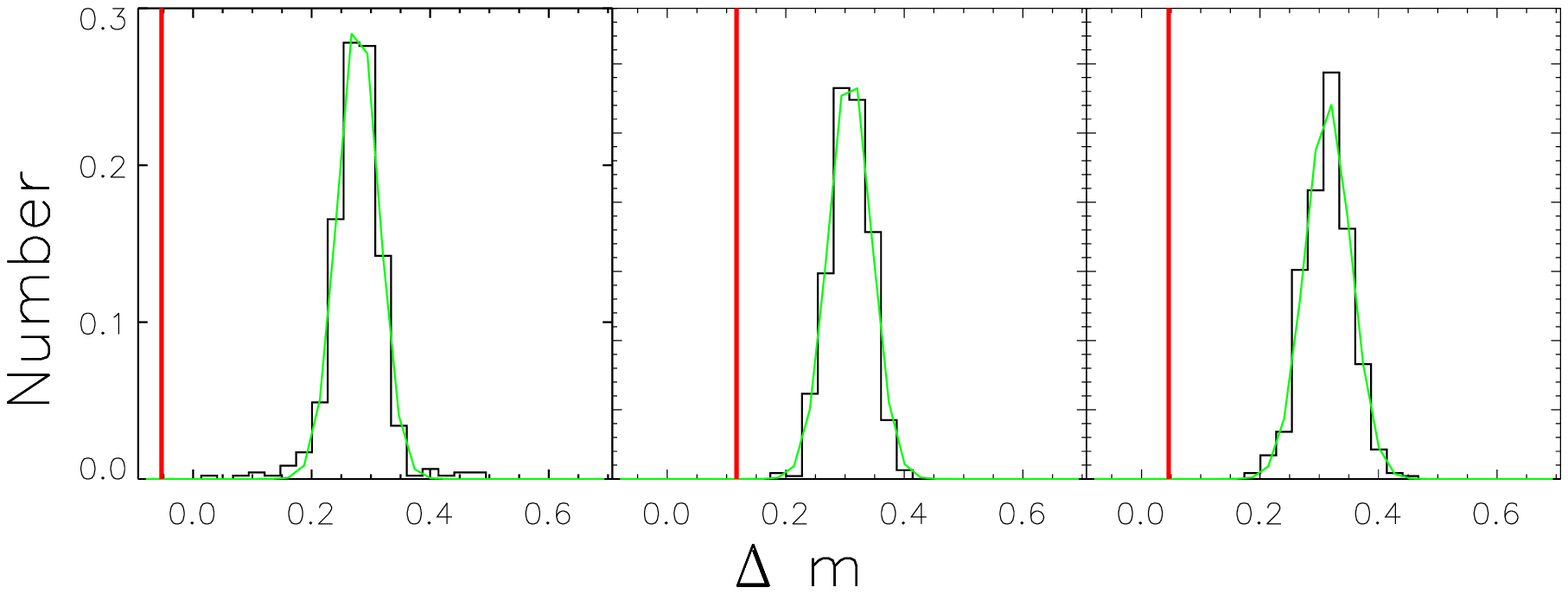}
\caption{Distributions of the difference between the fiber and spectroscopic magnitudes in $g$, $r$, and
$i$ bands (from left to right) for galaxies on the same plate of the ECLEs. For a comparison, the
differential magnitudes of the ECLE are indicated as a red line. The upper panels are for SDSS
J0748+4712 and the bottom panels for SDSS J0952+2143.}
\label{fig3}
\end{figure}

\begin{figure}
\plotone{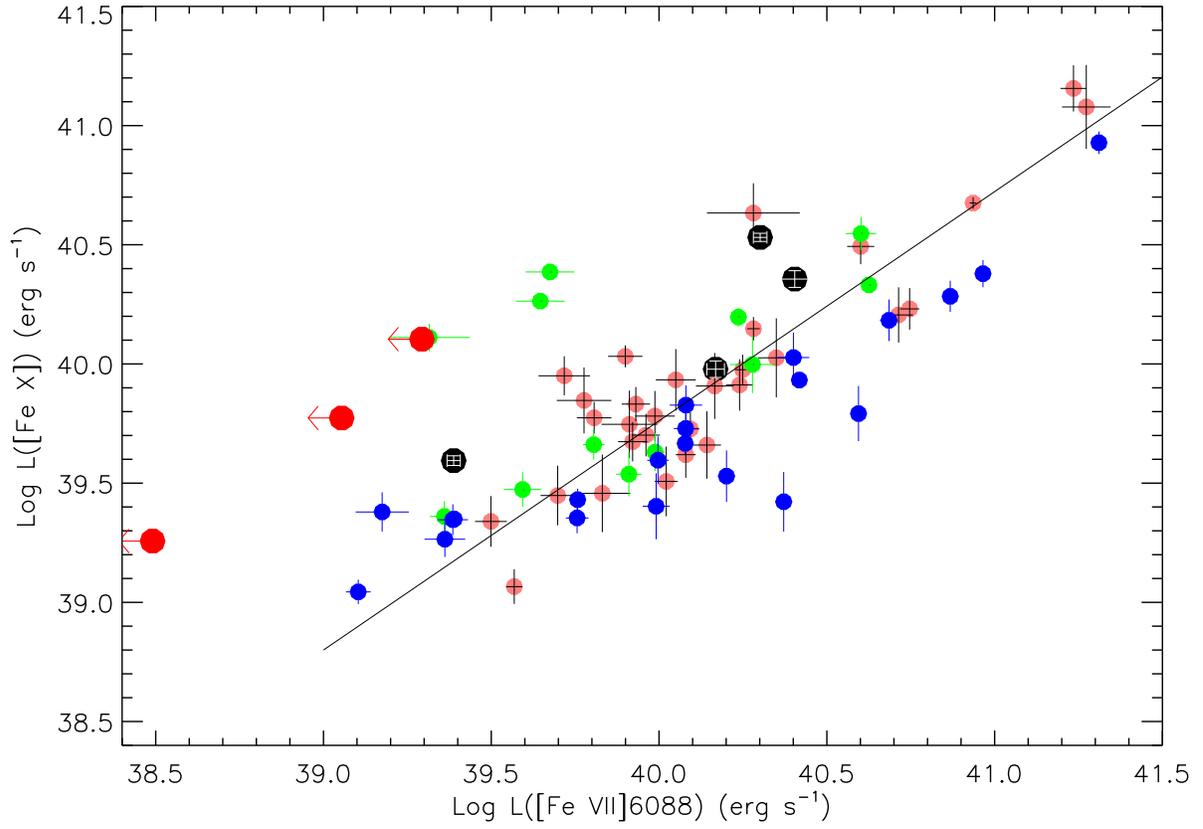}
\caption{\fex\ luminosity vs. \fevii$\lambda$6088 luminosity for the ECLEs. For a comparison,
Seyfert galaxies from the sample of Gelbord et al. (2009) are also shown. The color codes are:
broad-line Seyfert galaxies (pink), narrow-line Seyfert 1 galaxies (green), Seyfert 2 galaxies (blue), and ECLEs with (black) and without (red) \fevii. }
\label{fig4}
\end{figure}

It is evident that our ECLEs can be divided into two subclasses, with and without the detected
\fevii\ lines (Figure \ref{fig1} and Table \ref{tab2}). As shown in Figure \ref{fig4}, the 
lower limits on \fevii\ give \fex/\fevii$\lambda 6087>3$ for objects without \fevii\ lines, 
while those with detected \fevii\ have \fex/\fevii$\lambda 6087<3$. In Figure \ref{fig4}, we plot the 
\fex\ versus \fevii$\lambda$6087 luminosities for our ECLEs; for comparison those of a sample 
of Seyfert galaxies from Gelbord et al. (2009) are also overplotted. The ECLEs with \fevii\ 
show \fex/\fevii$\lambda$6087 ratios higher than normal Seyfert galaxies, and similar to 
narrow line Seyfert 1 galaxies, which are believed to have black holes of relatively small masses accreting 
at near the Eddington rate. The objects without detected \fevii\ show much higher ratios. 

\begin{figure}
\plotone{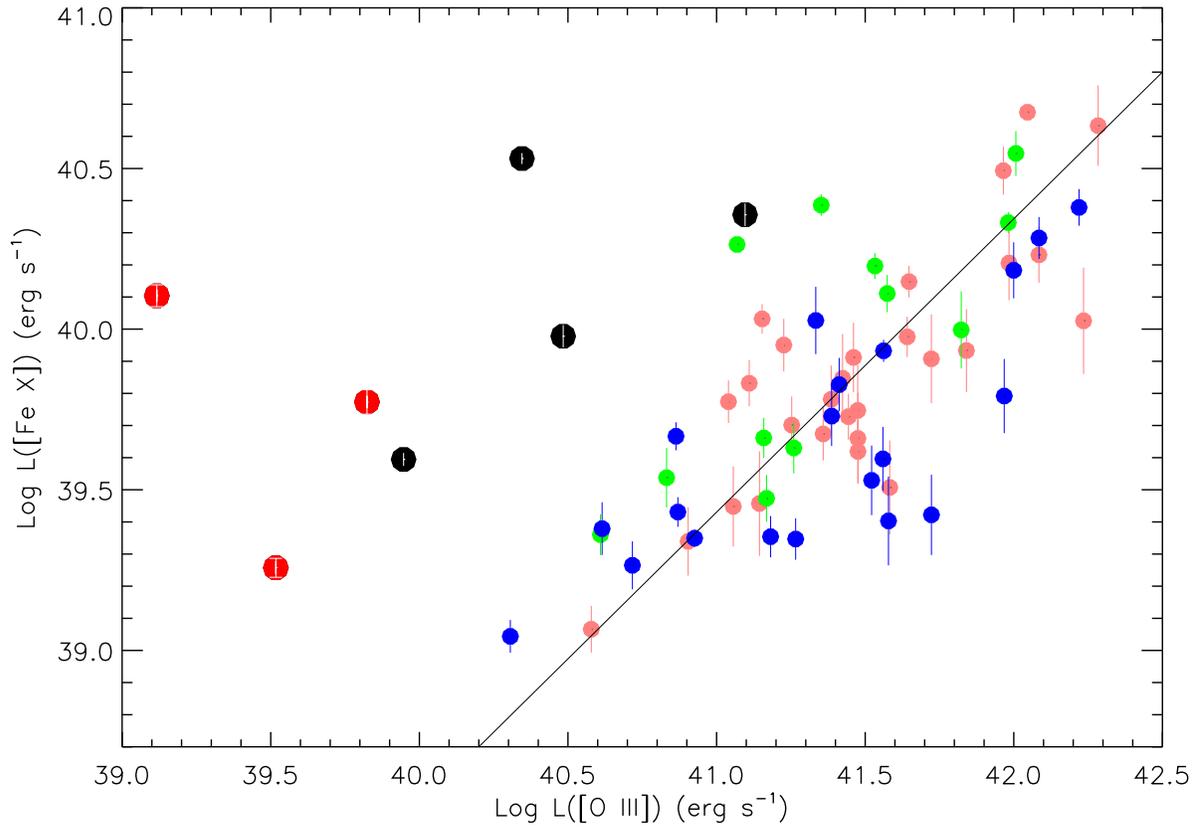}
\caption{\fex\ luminosity vs. \oiii\ luminosity for ECLEs and Seyfert galaxies (Gelbord et al. 2009). 
Color codes are the same as in Figure 4. }
\label{fig5}
\end{figure}

In Figure \ref{fig5}, we show the \fex\ versus \oiii\ luminosities. The \fex/\oiii\ ratios of 
the ECLEs are 1$-$2 orders of magnitude higher than those of Seyfert galaxies that is expected 
from our selection criteria. The ECLEs are outliers, offset from the smooth distribution of 
the Seyfert galaxies. There is no significant difference in the ratios between ECLEs 
with and without \fevii, although the latter show systematically lower \oiii\ luminosity.

\begin{figure}
\plotone{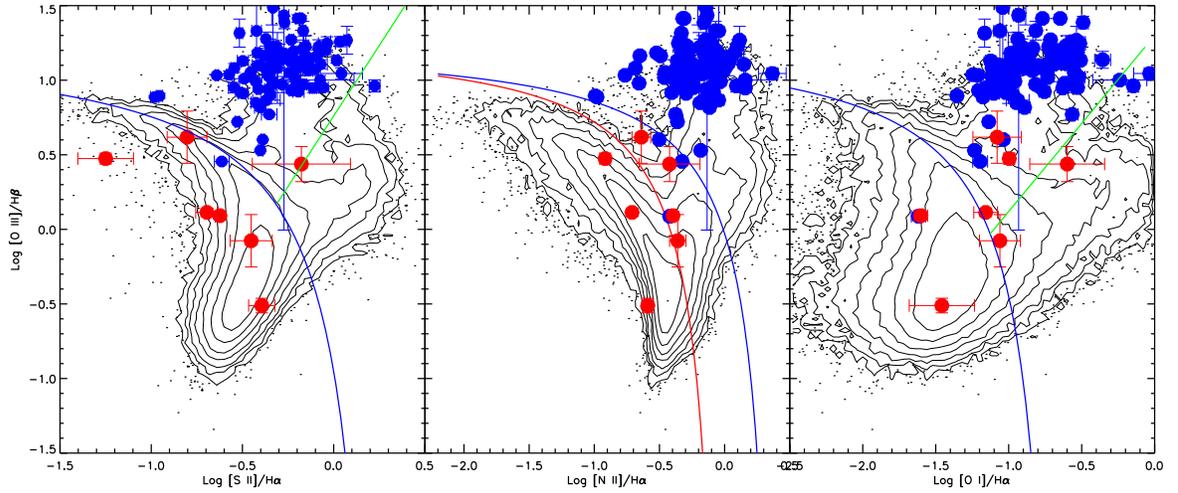}
\caption{Distribution of the ECLEs (red circles) on the BPT diagrams. Emission Line galaxies
from the SDSS are overplotted as number density contours. The blue lines represent the borders 
for extreme star-forming galaxies, the red lines represent the empirical boundary between 
star-forming galaxies and AGNs; the green lines separate LINERs and Seyfert galaxies (Kewley 
et al. 2006). For comparison, galaxies with coronal line detection in SDSS DR4 are plotted in blue.}
\label{fig6}
\end{figure}

To check whether they are actually Seyfert galaxies, we plot in Figure \ref{fig6} their 
locations in the BPT diagrams involving a set of narrow line ratios (Baldwin et al. 
1981; Veilleux \& Osterbrock 1987; Kewley et al. 2006) which are diagnostics 
of the ionizing continuum. For comparison, overlaid are the contours of the distributions 
of narrow emission line galaxies in the SDSS DR4 (Xiao et al. 2012). It can be seen that 
six of the seven objects lie within the \hii\ locus on the [S {\sc II}]/H$\alpha$ versus 
\oiii/H$\beta$ diagram, and the remaining one (SDSS J0952+2143) borders
LINER and \hii\ regions. On the [N {\sc II}]/H$\alpha$ versus [O {\sc III}]/H$\beta$ diagram, 
three (SDSS J0748+4712, SDSS J1241+4426, and SDSS J1341+0530) locate within the \hii\ 
(star-forming galaxies) locus, while the other four objects locate in the near \hii\ end 
of the composite region. However, among those four, the line ratios of three objects are 
indistinguishable from being \hii\ galaxies due to their relative large uncertainties. 
For comparison, other 
galaxies with reliable detections of CLs in SDSS DR4 spectroscopic 
data locate usually far away from the demarcating line (blue points in 
Figure \ref{fig6}; T.G. Wang et al., in preparation).
These results suggest that the bulk of our ECLEs differs largely from Seyfert galaxies, 
and the conventional narrow line regions in their galactic nuclei did not see
the hard ionizing continuum radiation that produces the strong CLs. The classifications appear
to be diverse when the \oi/\ha\ versus \oiii/\hb\ diagram is considered, where three objects
(SDSS J1055+5637, SDSS J1241+4426, and SDSS J0952+2143) locate in the AGN locus and one (SDSS J0938+1353) at
the demarcation of \hii\ and Seyfert galaxies. We note that the \oi\ line is, compared to other
narrow lines (e.g. [O {\sc III}], [N {\sc II}], [S {\sc II}] and \ha), generally broader 
and more difficult to accurately measure due to its weakness. In fact, given the high critical 
density of [O {\sc I}], it is possible that a significant part of the [O {\sc I}] emission is 
non-permanent; i.e., is contributed by the continuum flare which also produces the (variable) 
iron lines, and may come from a region co-spatial with the CL region. The 
\heii\ line is also prominent in the spectra of most of the ECLEs, suggesting that it 
is produced in the CL region.

\begin{figure}
\epsscale{0.92}
\plotone{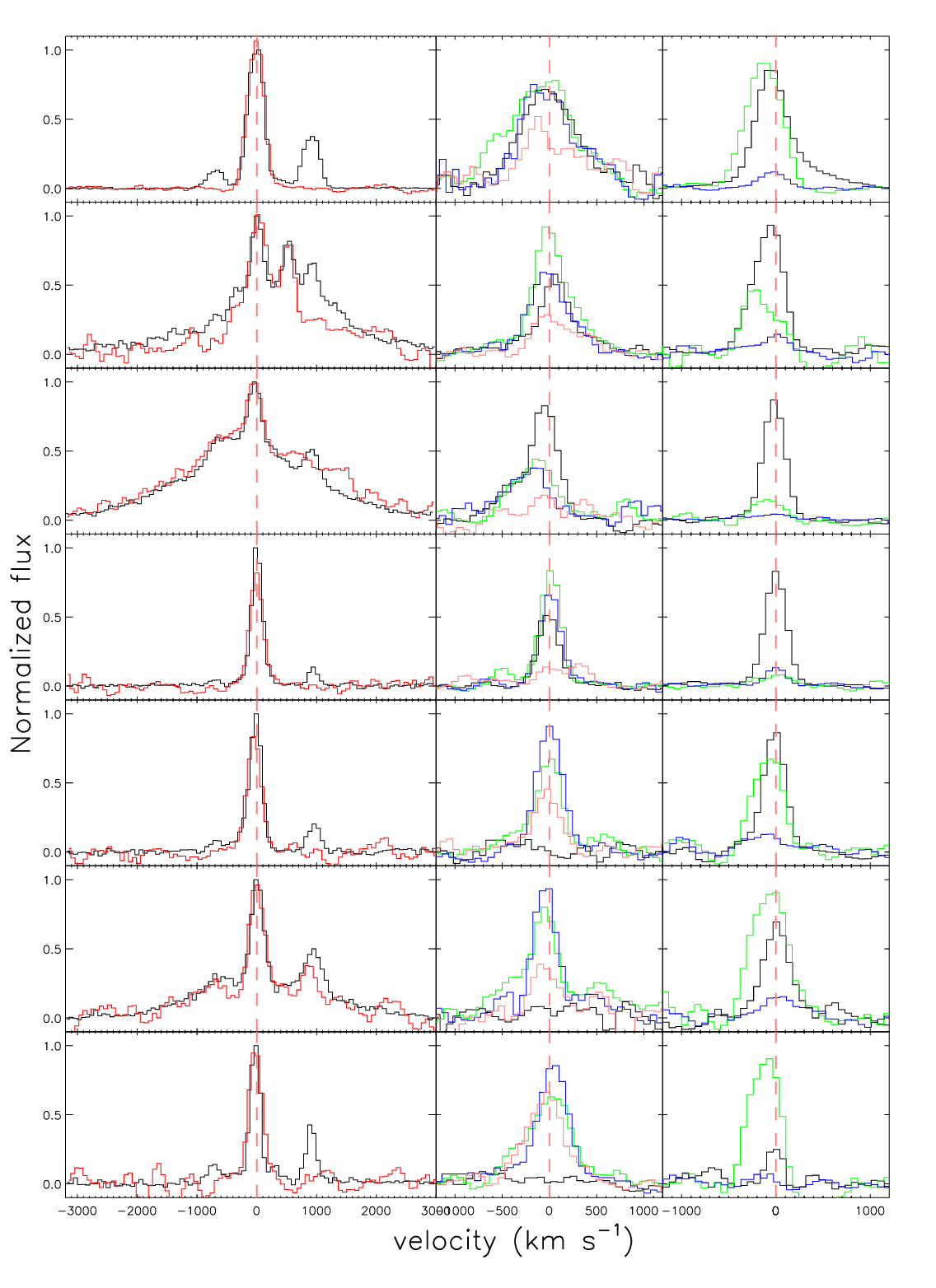}
\caption{Emission line profiles. The first column (from left to right): \ha(black) and \hb(red); the
second column: \fexiv\ (pink), \fexi\ (blue), \fex\ (green), and \fevii\ (black); the third column: 
\oiii\ (black), \oi\ (blue), and \oii\ (green). The objects are in the same order as in Figure 1.}
\label{fig7}
\end{figure}

\begin{figure}
\epsscale{0.9}
\plotone{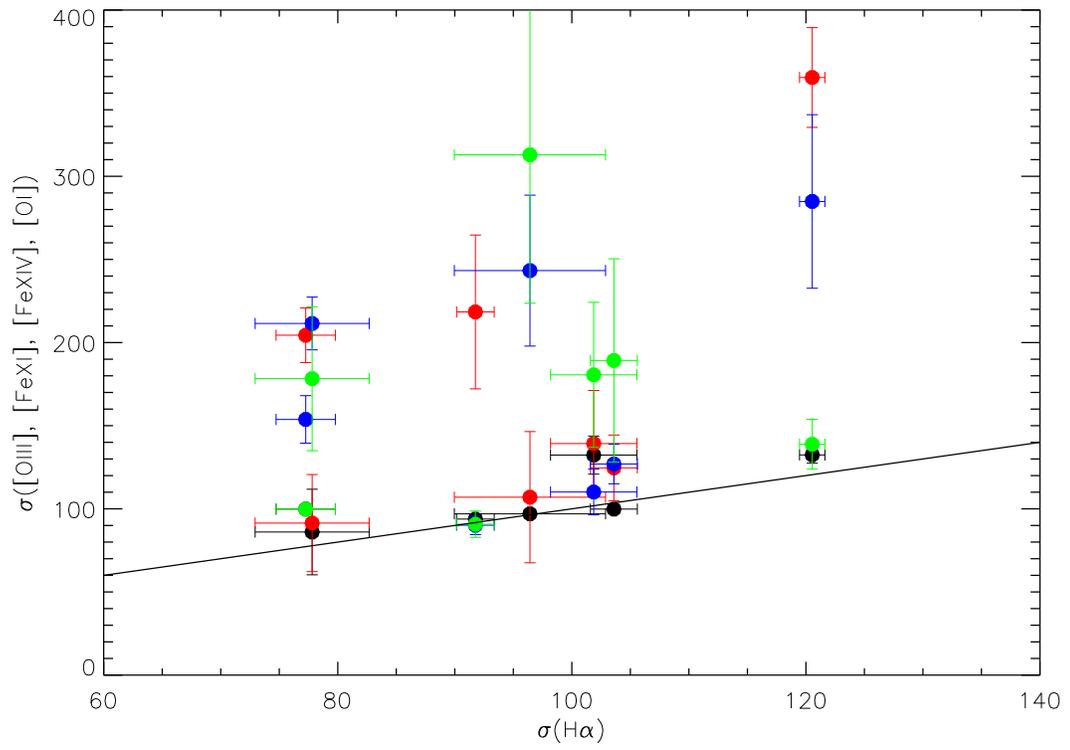}
\caption{Line width of the narrow component of \ha\ vs. the width of 
[O {\sc III}] (black), [Fe {\sc XI}] (blue), [Fe {\sc XIV}] (red) 
and [O {\sc I}] (green) line. The straight line denotes a one-to-one relation. 
}
\label{fig8}
\end{figure}

We also compare the profiles of the CLs with those of the other narrow lines (Figure \ref{fig7}).
The CLs usually display similar profiles, except that a narrow core is occasionally absent in high
ionization species, such as \fex, \fexi, and \fexiv. In most cases, the
CLs and [O {\sc I}] are broader than low-ionization lines (Figure \ref{fig8}). \heii\ 
has a similar profile to the CLs (Figure \ref{fig2}). For quantitative comparisons 
we calculate the FWHM from the best fitted
model. Note that some of the lines have apparently two components, and the FWHM may only
crudely reflect the kinematics of the line-emitting gas. The FWHM of \fex\ has a range
from 200 to 1000 km s$^{-1}$, and a median of 480 km s$^{-1}$. In comparison, \oiii\ has a median
FWHM of 230 km s$^{-1}$ and a range from 90 to 340 km s$^{-1}$, and \nii\ has similar values of 
a median 240 km s$^{-1}$ and a range from 110 to 340km s$^{-1}$. There appears to be no 
correlation of the line widths between the CLs and the low-ionization lines such as 
\oiii\ and \nii; however, the sample is too small for a meaningful test.

\subsection{Broad Emission Lines}

Broad emission lines are significantly detected in five out of the seven objects. For the three
objects with relatively high signal-to-noise ratio (S/N), the broad \ha\ and \hb\ lines have very similar profiles with 
complicated structures, such as wiggles and extra-peaks (left panels of Figure \ref{fig7}). It 
should be noted that extra-peaks are very rare among Seyfert galaxies, whereas they are present 
in two of the three objects in our study. One object is SDSS J0952+2143, which has already been 
presented in Komossa et al (2008, 2009). Another one, SDSS J1350+2916, shows a red peak and a 
blue peak at slightly larger velocities than SDSS J0952+2143. Due to their weakness, the broad 
\heii\ lines are detected only in SDSS J0938+1353, SDSS J0952+2143, and SDSS J1350+2916. The broad 
\ha\ lines have a median FWHM around 1400km s$^{-1}$ and a range from 880 to 2600 km s$^{-1}$. They are
much broader than the CLs, suggesting a different emission region. Very broad bumps are also
detected in J0748+4712; which were argued by Wang et al. (2011) as being attributed also to the
\heii\ and Balmer lines. The unusually strong \heii\ is probably due to the high helium
abundance in the debris of an evolved star that is tidally disrupted. In the follow-up observations carried out so far, these broad lines either disappeared completely or became considerably weaker. 
The strong fading and profile wiggle suggest that they are not of AGN origin.

\subsection{Host Galaxy Properties}

\begin{figure}
\plotone{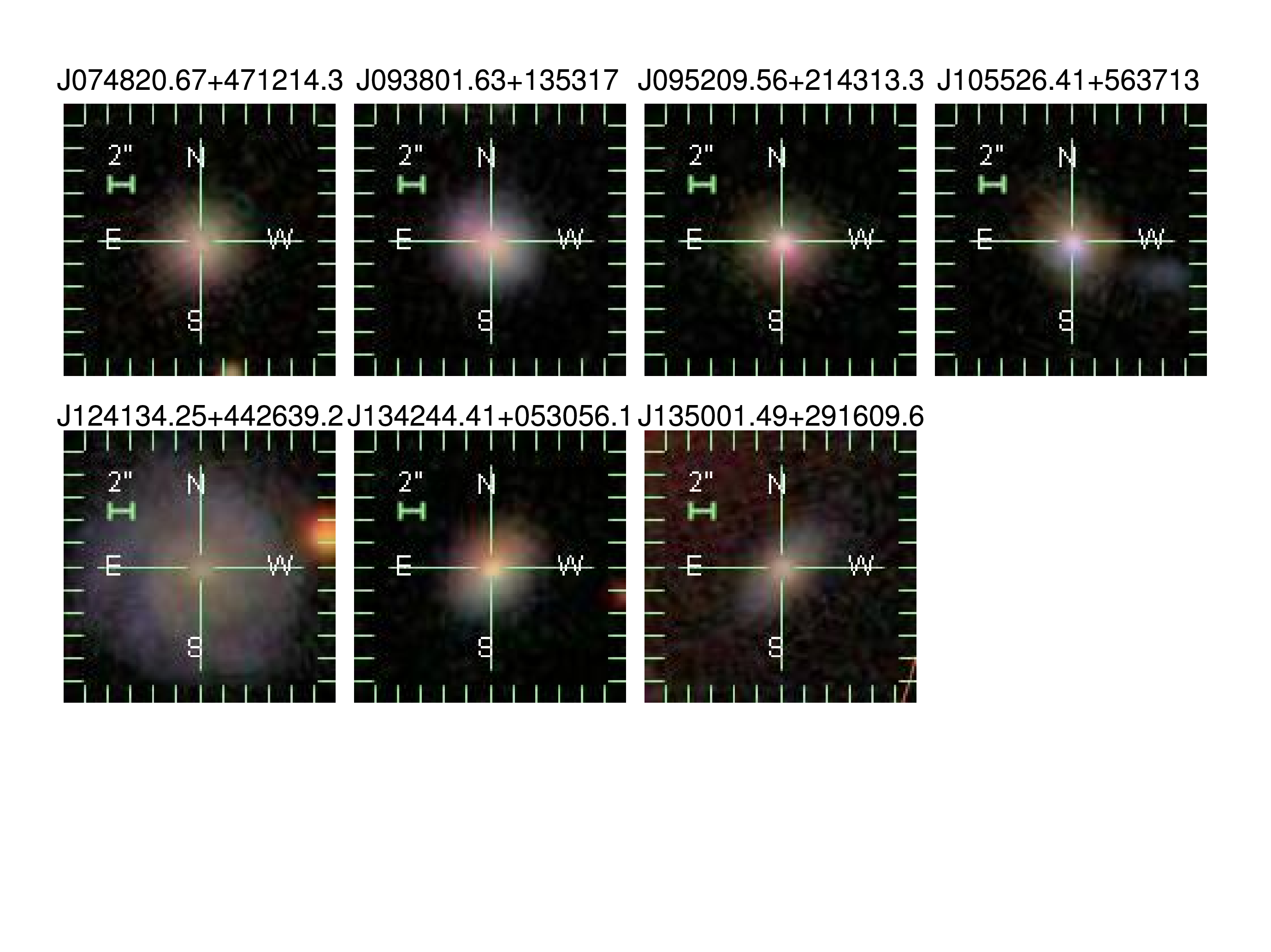}
\caption{SDSS images of the host galaxies in composite color.}
\label{fig9}
\end{figure}

Figure \ref{fig9} shows the SDSS images of the ECLEs. Most of these appear to be disk-galaxies
with the Petrosian radii in the range of 3$-$5 arcsec, except for SDSS J1241+2240, which shows
a relatively large low surface brightness disk. The Petrosian magnitudes are brighter than the
point spread function magnitudes by more than 2 mag in all these galaxies. Therefore, we estimate the absolute
magnitude of the host galaxies using the Petrosian magnitudes. The absolute magnitudes in the 
$i$ band are in the range of $-21.3 < M_i < -18.8$, suggesting that they are sub-$L_*$ disk galaxies in the
local universe (e.g., Blanton et al. 2003; Shao et al. 2007).

The stellar velocity dispersions, measured by fitting the starlight templates to the continuum
(Section 2), are mostly below 70 km s$^{-1}$, consistent with marginally resolved or unresolved absorption
lines. While all are small, we did not perform precise measurements, since they are close to (or
even below) the spectral resolution of SDSS (69 km s$^{-1}$) and the templates used are derived from a
stellar population library of low spectral resolution. The small stellar velocity dispersions are also
consistent with the narrowness of the [N {\sc II}] lines: $\sigma_{\rm line} < 40$ km s$^{-1}$ in four objects and only one
exceeding 100 km s$^{-1}$ albeit the large scatter in the $\sigma_*-\sigma_{\rm [N II]}$ relation (Greene \& Ho 2005).

\subsection{Spectroscopic Follow-up Observations}

Spectroscopic follow-ups have been carried out for three objects of our sample, and the high
ionization CLs were found to be fading ubiquitously. In the case of SDSS J0952+2143, the lines
of \fex, \fexi\ and \fexiv\ had decreased dramatically several years after
the initial SDSS spectroscopic observation. In contrast, the low-ionization lines [Fe VII] remained
nearly constant and \oiii\ even brightened in the very last observation (Komossa et al.
2009; H.Y. Zhou et al. 2012, in preparation). The broad lines decreased with time as well. A similar
trend was also observed in SDSS J1241+4426 in an MMT spectrum taken four years after the SDSS
observation (H. Y. Zhou, et al. 2012, in preparation). In SDSS J0748+4712, all the CLs disappeared
while \oiii\ increased by a factor of 10 in spectra taken 4-5 years later (Wang et al. 2011).
To summarize, in all the cases with follow-up observations so far, the high-ionization CLs and the
broad lines were fading on timescales of several years. The ionization of the line spectrum decreases
with time.

\section{The Nature of Extreme Coronal Line Emitters}

\subsection{Photoionization versus Collisional Ionization}

\begin{figure}
\plotone{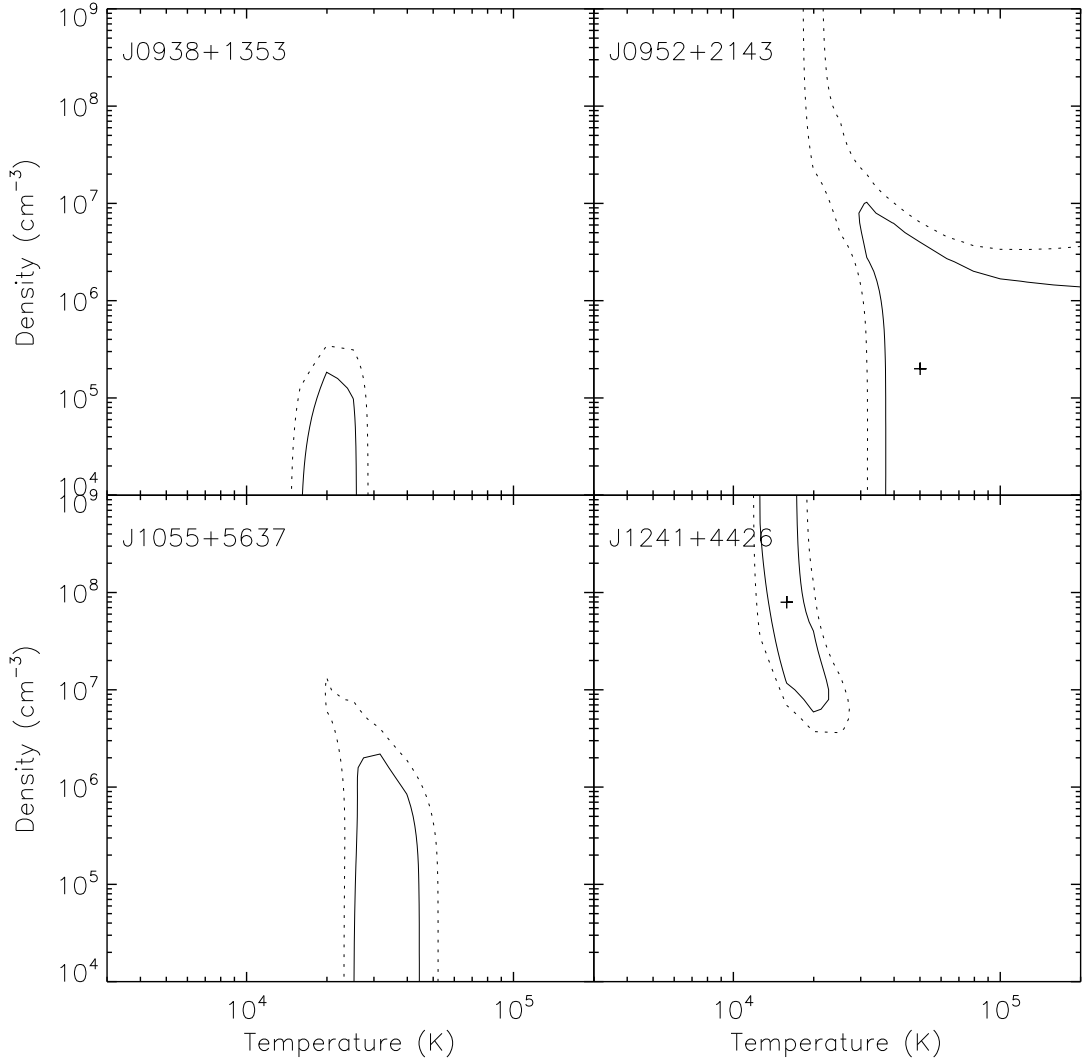}
\caption{Contours of the gas density and temperature inferred from the \fevii\ 
line ratios for four ECLEs with \fevii\ (SDSS J0938+1353, SDSS J0952+2143, SDSS J1055+5637, and 
SDSS J1241+4426). The solid and dashed lines denote contours at 68\% and 90\% confidence 
levels, respectively.}
\label{fig10}
\end{figure}

In this section, we show that the \fevii\ cannot be collisionally ionized, and argue that 
it is photoionized.
For objects with detected \fevii, the ratios of the \fevii\ lines can be used to constrain the
electron temperature and density. The \fevii5160/\fevii6088 ratio is from 0.1  to 0.3, while
the \fevii3759/\fevii6088 ratio is within 0.5 and 1.0, suggesting a range of the temperature
and density. We calculate the theoretical line ratios for a grid of temperatures and densities using
the atomic database CHIANTI (Dere et al. 2009), which are used to constrain the temperature
and density of the CL region in each object. The $\chi^2$ at each grid is computed for the observed 
line ratios and their uncertainties. In Figure \ref{fig10}, we show the confidence contours 
at the 68\% ($\Delta\chi^2=2.7$) and 90\% ($\Delta\chi^2=4.6$) levels in the density 
and temperature plane. 

In three of the four objects, the gas temperature is constrained to be 
within a few 10$^4$ K to 10$^5$K, while it is poorly constrained in the remaining object. 
These temperatures are less than 
what is required for collisional ionization ($T > 10^6$ K), thus photoionization is favored. 
An upper limit on the gas density is set to be $10^7$ cm$^{-3}$ for three objects, while a lower 
limit of $10^7$ cm$^{-3}$ is set for the other. Note that these ratios are not sensitive to the density 
below 10$^5$ cm$^{-3}$ or above $10^8$ cm$^{-3}$.
Obviously, spectra with higher S/N are required 
for better constraints on the density and temperature.

Lower limits on the line emissivity, defined as $n_en_{\rm ion}V$, are estimated for different ions using
the collisional strengths from CHIANTI, assuming that collisional de-excitation is not important.
In objects with detected \fevii, because \fex\ and \fexi\ have the critical densities
one order of magnitude higher than \fevii, they are less likely affected by collisional de-excitation.
For objects without \fevii\ detection, there is no good temperature and density diagnostics in the
optical band. In this case, we assume a gas temperature of $10^5$ K. Adopting a high gas temperature
will result in a higher emissivity. Assuming these ions are dominant species and the solar abundance,
we can estimate a lower limit on the total emissivity $\int n_en_{\rm H} dV$. These numbers are in the range
of $(5-50)\times 10^{60}$ cm$^{-3}$ (Table \ref{tab2}). Combing the line emissivity with the density, one can 
estimate the mass of the line-emitting gas. The typical value is from a few tenths to several solar 
masses, assuming a gas density of $10^4$ cm$^{-3}$.

\subsection{Constraints on the Soft X-Ray Energy}

The ions responsible for the CLs are created by ionizing photons in the soft X-ray band, ranging
from energies $>$125 eV for \fevii, $>$361 eV for [Fe {\sc XIV}], and to $>$650 eV for [Ar {\sc XIV}]. Therefore,
the soft X-ray luminosity incident on the CL region can be estimated with a detailed photoionization
model. Instead, in this paper, we will use the observed relation between the soft X-ray emission 
and the CLs in Seyfert galaxies to infer the soft X-ray luminosity, and leave detailed photoionization 
modeling to future work.

It was found that the fluxes of \fex\ and \fexi\ are well correlated with the soft X-ray flux 
in Seyfert galaxies \footnote{\fevii\ flux is not well correlated with soft X-ray flux.}
(Gelbord et al 2009). The CLs to X-ray flux ratios are $\log(f_{[{\rm Fe {\sc X}}]}/f_X)=
-3.43\pm0.55$ and $\log(f_{[{\rm Fe {\sc XI}}]}/f_{\rm X})= -3.52\pm0.38$ for both broad and narrow line Seyfert 
1 galaxies. These lines are thought to form in the inner edge of a dust torus (Murayama \& Tanigichi 1998a, 1998b; Rodr\'iguez-Ardila et al. 2002), which has a covering factor about 0.6$-$0.7 based on the 
fraction of type 2 Seyfert galaxies. As shown in Figure 3, the observed \fex\ luminosities of 
the ECLEs are in the range of 39.2 (dex)$-$40.5 (dex) in erg s$^{-1}$, similar to those of the 
CLs detected in AGNs. With the above flux ratios, we estimate the soft X-ray luminosities in 
the range of 42.5 (dex)$-$43.9 (dex) erg s$^{-1}$. Soft X-rays with a luminosity of $10^{41}$ 
erg s$^{-1}$ (0.1$-$10 keV) were detected in SDSS J0952+2143 in 2009 (Komossa et al. 2009). Since 
the emission line fluxes were almost as high as the quasi-simultaneously measured X-ray flux, 
the authors suggested that the lines must be the echo of a past X-ray flare with a much higher 
luminosity. We note in passing that objects in Gelbord et al (2009) are classical Seyfert galaxies, which 
are persistent sources; and so they are not affected by the fading effects seen in these ECLEs.

To estimate how much energy in soft X-rays is required, we integrate the CL energy 
over time. Assuming a lifetime for \fex\ and \fexi\ of one year, the integrated 
\fex\ or \fexi\ line energy is in the range of 46.7 (dex) $-$48.0 (dex) erg. With the 
above $L($\fex$)/L_X$ and $L($\fexi$)/L_X$, the total energy in soft X-rays is 
estimated to be around 50 (dex) $-$51.4 (dex) erg.

\subsection{Event Rate for ECLEs}

We estimate the event rate from the number of ECLEs, the lifetime of CLs and the number
of galaxies observed. From the follow-up observations of the three objects, we find that 
extreme CLs are visible for 3$-$5 years. This gives a lifetime of the order of three  
years. We found seven ECLEs from 676,881 galaxies and quasars at redshifts less than 0.20 
in the SDSS DR7 spectroscopic sample. This gives an event rate of 2$\times10^{-6}$ 
galaxy$^{-1}$ yr$^{-1}$. 

The event rate is dependent on galaxy luminosity. As mentioned above, 
the host galaxy luminosities of the ECLEs are clustered in a relatively narrow 
luminosity range ($-21.5 < M_i <-18.5$, see Figure \ref{fig10}). This dependence may 
be related to the physical conditions which are required to produce the CLs observed, 
such as the central black hole mass or the gas environment. Furthermore, in practice 
there is a lower limit on the spectral S/N ratio for the detection of CLs. Considering 
this effect, we set a lower S/N cutoff to the spectra as the lowest spectral S/N ratio 
(S/N=14.8 in the $i$ band) among the detected ECLEs. As a result, we obtain an 
event rate of about 1.5$\times 10^{-5}$ per galaxy per year for all the galaxies with 
the absolute magnitudes in the range of $-21.5 < M_i < -18.5$. Despite a large number 
of more luminous galaxies that meet the S/N ratio constraints, we did not detect any 
ECLEs among them.

\subsection{Difference between ECLEs with and without \fevii}
 
We do not detect \fevii\ in three of the seven objects. As clearly seen in both Figure \ref{fig4} and
Table \ref{tab2}, the non-detection is not due to the low S/N of the spectra, but due to the weakness
or absence of the \fevii\ lines. The \fevii\ lines have significantly lower critical densities
($10^6-10^7$ cm$^{-3}$) and lower ionization potentials than \fex, \fexi\ and \fexiv.
The lack of \fevii\ in the three objects may be due to one or both of the following two factors:
\fevii\ is collisionally de-excited in high density gas, while \fex\ and \fexi\ are not
because of their higher critical densities; the gas is overionized under intense soft X-ray radiation.
In Section 3.1, we showed that in one object with \fevii\ detection the density of the CL region is close to the critical
density of \fevii. Therefore, it is possible that some objects may have the densities even higher
so that \fevii\ is collisionally de-excited. On the other hand, it is also plausible that objects
without \fevii\ have higher ionization parameters. All the three objects without \fevii\ show
the \sxii\ lines, while only two of the four objects with \fevii\ do. However, we cannot
verify this as there is no good density indicator in the optical band for objects without \fevii\
lines. Below, we investigate several possibilities in order to understand the origin of this difference.

We examined other line ratios, \fexiv/\fexi, \oiii/\fexi, and
\heii$_n$/\hb$_n$ and find no significant difference between objects with and without the \fevii\
lines. The intensity of  \fexiv\ is comparable to those of \fex\ and \fexi\ in
both subclasses. However, the \oiii\ luminosities for objects without \fevii\ are one order
of magnitude lower than those with \fevii. The host galaxies of ECLEs with the \fevii\ lines
appear to be more luminous than those without. All the host galaxies of the ECLEs with \fevii\ 
are brighter than -19.6 mag in the $i$ band, whereas those without are fainter than this value. Among
the four objects showing continuum variations between the SDSS photometric and spectroscopic
observations, the two with \fevii\ have higher non-stellar luminosities ($M_{g,{\rm var}} < -17.5$) than the
two without \fevii\ ($M_{g,{\rm var}} > -17.0$), as seen in Table 1. However, these results should be tested
with a larger sample in the future.

\subsection{Tidal Disruption, Nuclear Activity or Peculiar Supernova}

The strong CLs clearly require a strong soft X-ray ionizing source. In galactic nuclei, a number
of processes may produce strong soft X-ray emission: an SN explosion, episodic accretion onto a
supermassive black hole following a TDE or disk instability, or encountering a molecular cloud
(Komossa et al. 2008, 2009; Wang et al. 2011). Given the large mass and size of a stable molecular
cloud, events of random encounter of a molecular cloud likely last for a much longer timescale.
The fact that most of the objects avoid the AGN region in the BPT diagram suggests that the
normal narrow line region did not see the hard ionizing continuum from the accretion disk. This
implies that the last nuclear activity was either at least several $10^3$ years ago (the light 
crossing time of traditional narrow line region) or at a very low level. The decay timescales are much shorter 
than those of limit-cycle outbursts driven by local thermal instability (Lin \& Shield 1986) 
in accretion disks. Thus, episodic feeding by a molecular cloud, a persistent AGN and disk 
instability is very unlikely.

CLs are observed in a few SNe of type IIn, mostly at a late evolutionary stage. They are
produced in pre-shocked regions ionized by radiative shocks from an expanding shell running 
into clumpy circumstellar medium (Smith et al. 2009). However, the CL luminosities in these SNe
are factors of tens to hundreds lower than observed in the ECLEs studied here. Furthermore, the
spectra of these SNe also display rich, strong low-ionization lines such as Fe {\sc II} at late 
stages, or strong He {\sc I} emission lines in the early spectrum of SN 2005ip (Smith et al. 2009); 
these features are very different from the ECLE spectra. In addition, the current models of SNe do 
not predict sufficient energy in soft X-rays either. Nakar \& Sari (2010) calculated the X-ray 
light-curves of core-collapsed SN breakouts for various initial masses. Integrating these 
light-curves yields a total soft X-ray energy about one to several orders of magnitude lower than 
those estimated in the last section. The X-rays produced in late shocks are also one order of magnitude 
lower than the those required to power the CL luminosity, even in strongly interacting 
objects (Fabian \& Terlevich 1996; Chevalier \& Fransson 1994). Finally, as discussed in Section 3.1, 
the UV magnitude of SDSS J0748+4712 measured 20 days after the detection of the CLs was brighter than 
$-$17 mag, which is too bright for any current SN model predictions (e.g., Nakar \& Sari 2010). 
For these reasons, we consider SNe to be unlikely either.

Most of the observed properties can be reconciled in the context of tidal disruption of stars by
massive black holes at galactic centers (Komossa et al. 2009). Fallback of the bounded tidal debris
forms an optically thick accretion torus around the black hole. The luminosity is peaked at near
or super-Eddington on the accretion timescale, and then decreases as $t^{-5/3}$ (Rees 1988; Evan \&
Kochanek 1989; cf. Lodato et al. 2009). If ∼10\% of the radiation goes into soft X-rays,
it gives an integrated soft X-ray energy of $\sim 0.01M_*c^2 \simeq 10^{52}$ erg, which is 
sufficient to power the CLs, assuming 10\% radiation efficiency in the 
accretion disk. A TDE can explain naturally (e.g., Strubbe \& Quataert 2009) 
the bright UV source detected in SDSS 
J0748+4712 at least 20 days after the CL detection.

Broad emission lines can be produced via reprocessing the disk emission by bound and unbound
stellar debris (Eracleous et al. 1995; Bogdanov\'ic et al. 2004; Komossa et al. 2008; Strubbe
\& Quataert 2009). Blueshifted, very broad lines, seen in SDSS J0748+4712, can form in winds
driven by super-Eddington accretion during the initial flare (Strubbe \& Quataert 2009). As in
SNe, unbound material with high velocities may interact with the surrounding ISM, and produce
broad emission lines as well (e.g., Khokhlov \& Melia 1996; Ayal et al. 2000). In that case, 
material emitting the Balmer lines can be collisionally excited, rather than photoionized. 
The debris has a range of orbits and may be dynamically unsettled. This produces variable 
broad lines with complex profiles, as seen in the observed spectra. As a consistence check, 
the mass of the line-emitting gas should be smaller than that of the disrupted debris. 
Assuming the broad Balmer lines are produced via recombination, the minimum mass of the \hii\ 
region can be written as $M_{\rm min}=1.3m_{\rm H}N_pL_{{\rm H}\alpha}/(4\pi j_{{\rm H}\alpha,br})\simeq 0.03 
n_{e,10}^{-1} L_{{\rm H}\alpha,br,41}$ $M_\sun$ using the H$\alpha$ line emissivity from 
Osterbrock \& Ferland (2006) for a temperature $T=$10,000 K, where $n_{e,10}$ is the 
electron density in units of $10^{10}$ cm$^{-3}$. The observed $L_{{\rm H}\alpha,br,41}$ luminosities 
(in units of $10^{41}$ erg~s$^{-1}$) are in the range of 0.03$-$3.0, which are consistent with 
stellar debris of a few tenths solar mass as line-emitting gas for a reasonable gas density. 

CLs may come from interstellar medium photoionized by a flare. Typically a flare lasts for
a few months, which is longer than the recombination time for the gas density inferred from 
the \fevii\ line ratios.\footnote{At a temperature of $10^5$K, the recombination time is 
5$\times10^4(n_e/10^6$~cm$^{-3}$) s for Fe$^{+6}$, and 20 and 30 times smaller for Fe$^{+9}$ 
and Fe$^{+10}$.} At any given time, emission lines come from a region within two parabolic 
iso-delay surfaces of a delay interval of months at large scales as well as from the nuclear 
region (Figure \ref{fig12}). The ionizing photons, i.e., soft X-rays, must reach a distance 
a few light-years in order to explain the presence of the CLs for a few years. In the last 
section, we have shown that there is a quite large range of densities, from object 
to object, from a few 10$^4$ cm$^{-3}$ to $>$ a few $10^7$ cm$^{-3}$ for the ECLEs with \fevii. 
If the line-emitting gas fills completely the region, it would be optically thick to the soft 
X-rays. Thus, the emission line region must be clumpy. Only the dense clouds produce the CLs 
and some recombination lines such as the Balmer lines and He {\sc II} line, while the low 
density gas is probably completely ionized in the strong radiation field of the flare and is
transparent to the soft X-rays.

Tidal disruption may leave significant imprints on the kinematics of the line-emitting gas. 
Gas exposed to the intense radiation will receive a radial acceleration, and over the period 
of the passage of the flare, gas will gain a radial velocity of the order $k (\Delta 
t/{\rm 1\;month}) (t/{\rm 1\; month})^{-2}$ \kms, where $k=[(\kappa/\sigma_T) (L/L_{\rm Edd})-1]$ 
is the ratio between the radiation force and gravity, and $\Delta t$ the duration of the UV/X-ray 
flare. $k$ can be on the order of $10^2$, even if the gas contains a small amount of dust. This may 
explain the blueshifts seen in some of the CLs. The radial velocity decreases rapidly with time, 
thus we expect that the blueshifts become smaller as the flare evolves. Also the line width becomes 
smaller with time as gas further out contributes to the CLs. However, the distribution of the line
emitting gas is likely not affected much by TDEs. This is because the timescale of density adjustment
is the sound-traveling time within individual clouds, and the timescale for cloud migration is $r/v$,
where $v \sim 10^2$km s$^{-1}$ is estimated from the emission line widths.
A decrease of ionization with time was observed in three objects analyzed so far. Such a
behavior is expected in a number of cases: gas density falls less steep than $r^{-2}$, so the ionization
parameter decreases as the emission line region moves out; if the optical depth to soft X-ray
absorption is moderate, the ionizing continuum may be diluted on the way outward. In addition, as
the mass accretion rate decreases with time, the ionization parameter in the inner region decreases 
and the continuum may be softened, too. Detailed photoionization modeling is needed to understand
this problem.

If these ECLEs are indeed associated with TDEs, the question arises as to whether any TDEs,
regardless whether discovered in X-ray, UV, or optical surveys, show strong CLs. To
answer this question, we searched for post-flare spectra of the TDE candidates reported in the
literature. The optical spectra of the {\em ROSAT} TDE candidates were mostly taken 5-10 years after
the flares, and no high-ionization emission lines were detected in dedicated searches 
\footnote{The one exception is the AGN IC3599, which underwent a large soft X-ray outburst and 
had an optical spectrum taken within one year. The high-state optical spectrum showed strong 
CLs of [Fe {\sc XI}] and [Fe {\sc XIV}], which faded in subsequent years, plus a broad 
Balmer emission line of 1200 km~s$^{-1}$ width which also faded (Brandt et al.
1995; Grupe et al. 1995; Komossa \& Bade 1999). IC3599 had been known to be an AGN before the 
X-ray flare.} 
(Komossa \& Bade 1999; Komossa \& Greiner 1999; Gezari et al. 2003). We found post-flare spectra within
less than five years of the flares for eight objects among the most recent candidates identified from X-ray,
UV, and optical surveys (Gezari et al. 2008, 2009; Esquej et al. 2008; van Velzen
et al. 2011; Cenko et al. 2012), but no strong CLEs were detected. These results suggest that
most TDE candidates identified so far do not show strong CLs. However, Gezari et al.
and Cenko et al. rejected objects either with a broad line or with \oiii/\hb$>$3 to remove
AGN, while most objects in our sample do show (possibly transient) broad lines and two have 
\oiii/\hb$> 3$. Therefore, by definition, they also excluded any potential TDEs with strong 
CLs. Furthermore, their absorption-line dominated TDEs do not produce CLs as discussed later.
We checked their rejected spectra. In fact, among eight objects with UV flares and
with optical follow-up spectra in their sample, we find that one object, D2-9 (J100002.0+024216),
displayed a very strong \fexiv\ line (see Figure 7 of Gezari et al. 2008). This object was classified
as Seyfert 2 based on \oiii/\hb$>3$ by those authors. If its \fexiv\ line turns out
to be fading in future observations, it would be one TDE candidate. For most of the other objects, however,
the spectra are of too low S/N or do not cover the CL regime. Thus, the fraction of
TDEs with strong CLs is not constrained.

To produce strong CLs, tidal disruption must occur in an environment of rich cold gas. Note
that all objects in our sample show narrow emission lines from a region different from the  CL 
region as indicated by their widths and/or low-ionization levels, demonstrating the presence
of cold gas. The non-detection of CLs in absorption-line galaxies appears to be a natural
outcome of depletion of cold gas in their nuclei. Whether or not a gas-rich galaxy (i.e., \hii-type
galaxy) does show strong CLs after a nuclear flare would also depend on the amount of gas near the
nucleus that is capable of producing CLs. In addition, it is also necessary to have strong soft X-ray
emission from the accretion disk to ensure extreme CL emission. While most high-energy
flares were indeed characterized by very luminous soft X-ray emission, some UV-detected flares may
have been weak in X-rays (Gezari et al. 2009), and the broad-band spectral energy distribution
depends on the details of the stellar debris evolution, and on possible effects of self-absorption.

As we discussed in Section 3.4, ECLEs are hosted in intermediate-luminosity disk galaxies. Besides,
these galaxies do not show apparent bulges in the SDSS image, indicating that the bulge luminosity 
would be much lower than that of the whole galaxy. This can be seen most clearly in the nearest object,
SDSS J1241+4426, where we found only a faint light concentration in the very center of the galaxy.
Thus the black hole mass in these galaxies is likely small according to the black hole and bulge mass
relation. A similar conclusion can be reached from the $M_{\rm BH}-\sigma_*$ relation (Tremaine et al. 2002). If we
use the stellar velocity dispersion in Section 3.4, the black hole masses would be at most around $10^6$ 
$M_\sun$. Because a high soft X-ray luminosity is required to power luminous CLs, the black hole mass is
not likely much below $10^5$ $M_\sun$.

Wang et al. (2011) speculated that ECLEs with and without \fevii\ are at different evolution
stages based on the variations of [Fe {\sc X}] and \fevii\ in three objects with follow-up 
observations so far. The high-ionization lines faded much faster than \fevii. Thus, we expected 
higher ionization in earlier stages. The relative fraction 
of each subclass depends on the time spent in each stage, i.e. the similar numbers detected 
for each subclass indicate similar time periods spent in each stage. However, this cannot explain 
their difference in the luminosity of host galaxies, in [O {\sc III}] luminosity and possibly in 
the luminosity of variable components, if confirmed.

Alternatively, these differences may be understood in terms of the differences in black hole
mass, and host galaxy properties. Objects with the \fevii\ emission lines appear to be systematically 
more massive than those without. Less massive galaxies tend to host smaller black holes, which may 
result in a hotter accretion torus and a harder soft X-ray spectrum (Montesinos Armijo \& de Freitas 
Pacheco 2011). This may lead to over-ionization of the line-emitting gas, and thus very weak \fevii. 
Furthermore, if more massive host galaxies have a different gas distribution in their cores (gas 
mass, density, and column density), it is conceivable that the \fevii\ emission would be affected 
in a systematic way.

\section{Conclusion}

We have compiled a sample of seven ECLEs, which have \fex/\oiii\ 
one order of magnitude higher than those of Seyfert galaxies from the SDSS spectroscopic
samples of galaxies and quasars. The luminosities of the CLs are on the order of 
$10^{40}$ erg s$^{-1}$, comparable to those observed in Seyfert galaxies. Despite the 
strong CLs, most of these galaxies locate in the regime of star-forming galaxies 
on the BPT diagrams, suggesting that the conventional narrow-line region did not see the 
hard ionizing continuum that the CL region did. Follow-up observations of three of 
the ECLEs show that the high-ionization CLs had been fading, while the low-
ionization lines remained constant or even brightened on timescales of several years. 
In four objects, variations of the continuum emission are detected between the SDSS
imaging and spectroscopic observations which spanned from 4 months to two years. These 
also set constraints on the ages of ECLEs from less than 3 months 
to more than two years. Broad emission lines are detected in five of the seven objects. 
These lines exhibit complex profiles, including blue and red peaks in two objects, 
and a very broad and blueshifted line in one. In three of the objects the broad emission 
lines diminished, similar to the high-ionization CLs, as revealed by follow-up 
observations. The ECLE sample can be divided into two subclasses of similar numbers, 
those with and without the detected \fevii\ emission. They do not show significant 
difference in the \fexiv\ to \fex\ ratios. The ECLEs with detected \fevii\ tend to have 
brighter host galaxies, higher \oiii\ and higher variable continuum luminosities
than those without. In general, their host galaxies are disky and underluminous, 
in the range of $-21.5 < M_i < -18.5$, and have small stellar velocity 
dispersions. Based on the current sample, we estimate an event rate for the ECLEs 
on the order of $\sim 10^{-5}$ galaxy$^{-1}$ yr$^{-1}$ for galaxies of 
$-21.5 < M_i < -18.5$.

Various scenarios are discussed to explain ECLEs. The most promising one is tidal 
disruption of a star by less massive black holes residing in cold-gas-rich galactic 
nuclei. In this scenario the broad lines result from the reprocessing of UV/X-ray flares 
by the unbound and/or bound stellar material, while the CLs are produced by the 
clumpy interstellar medium. The diverse properties of the emission lines may be understood 
qualitatively in terms of the differences in the gas environment, the shape of the ionizing 
continuum, and the stage of time evolution. Specifically, the dichotomy between having 
and lacking the \fevii\ lines may be ascribed to the dependence of soft X-ray spectrum on 
the black hole mass, or nuclear gas density. However, we still do not understand the 
key physical processes responsible for such a diversity. We are carrying out spectroscopic follow-ups 
of these objects in order to study long timescale variations and to further constrain their nature.

If these ECLEs are confirmed to be associated with TDEs, our approach may provide a new
method of a systematic search for TDEs from large spectroscopic surveys. The extreme CL phase
lasts for a few years, as observed in two objects, much longer than the photometric flaring 
signatures in the optical and UV. Spectroscopic detections and subsequent monitoring of the long term
evolution of emission lines can provide rich diagnostics on the ISM in the vicinity of a dormant black
hole, which can be hardly obtained otherwise. The event rate can be used to constrain the incidence
of supermassive black holes in sub-$L_*$ galaxies, when combined with the host galaxy properties. 
We plan to carry out such surveys with the Chinese LAMOST telescope to conduct early spectroscopic
observations, as well as multi-waveband follow-ups. Because most of these ECLEs show 0.1-0.3
mag photometric variations within a 3\farcs0 aperture, they can be easily picked up with the on-going
and future time-domain surveys such as PanSTARRs and LSST. Spectroscopic follow-up of such
flares in galactic nuclei may reveal a large population of ECLEs.

\acknowledgements
We thank the referee for helpful comments, which improved the content 
and clarity of the paper. T.W. thanks Gary Ferland for useful discussion 
on the photoionization models. This work was supported by the Chinese 
NSF through NSFC-10973013 and NSFC-11033007, the 
national 973 program 2007CB815403, and CAS knowledge innovation project 
No. 1730812341. 
This work has made use of the data obtained by SDSS. 
Funding for the SDSS and SDSS-II has been provided by the 
Alfred P. Sloan Foundation, the Participating Institutions, the
National Science Foundation, the U.S. Department of Energy,
the National Aeronautics and Space Administration, the Japanese
Monbukagakusho, the Max Planck Society, and the Higher Education
Funding Council for England. The SDSS Web site is http://www.sdss.org/.

\clearpage
\renewcommand{\thefootnote}{\alph{footnote}}
\begin{deluxetable}{cccccccccc}
\tabletypesize{\scriptsize}
\tablecaption{Basic Data of the Extreme Coronal Line Emitters \label{tab1}}
\tablewidth{0pt}
\tablehead{
\colhead{No} & \colhead{Name} & \colhead{Type$\tablenotemark{a}$} & \colhead{$z$} & \colhead{$M_{i,{\rm tot}}$} & \colhead{$R_{50}$}  
 & \colhead{$\Delta g\tablenotemark{b}$} & \colhead{$\Delta r\tablenotemark{b}$} & \colhead{$\Delta i\tablenotemark{b}$} & \colhead{$\Delta t\tablenotemark{c}$} \\
\colhead{} & \colhead{} & \colhead{} & \colhead{} & \colhead{(mag)} & \colhead{(arcsec)} & 
 \colhead{(mag)} &\colhead{(mag)} & \colhead{(mag)} & \colhead{(day)} 
}
\startdata
1 & SDSS J074820.66+471214.6 & No  & 0.0615&$-$19.75 & 4.88  & 0.23$\pm$0.04 & 0.09$\pm$0.03 & 0.13$\pm$0.04 & 120\\
2 & SDSS J093801.64+135317.0 & Yes & 0.1006&$-$21.29 & 3.77  & $<$0.22 & $<0.16$   & $<$0.14 & 326    \\
3 & SDSS J095209.56+214313.3 & Yes & 0.0789&$-$20.41 & 3.51  & $-$0.33$\pm$0.03& $-$0.19$\pm$0.04 & -0.27$\pm$0.04 & 375\\
4 & SDSS J105526.43+563713.3 & Yes & 0.0743&$-$20.01 & 3.71  & $-$0.20$\pm$0.07& $-$0.14$\pm$0.04 & -0.16$\pm$0.05 & 140\\
5 & SDSS J124134.26+442639.2 & Yes & 0.0419&$-$19.95 & 11.10 & $<$0.13 & $<0.12$ & $<$0.12  & 340  \\
6 & SDSS J134244.42+053056.1 & No  & 0.0366&$-$18.91 & 4.21  & $<$0.30 & $<$0.23 & $<$0.22  & 386 \\
7 & SDSS J135001.49+291609.7 & No  & 0.0777&$-$19.76 & 0.45  & 0.12$\pm$0.04 & $<$0.12& $<$0.13 & 701 \\
\enddata

\tablenotetext{a}{The type denotes whether \fevii\ is present or not.}
\tablenotetext{b}{Magnitude difference within the fiber aperture of 3 arcsec between the SDSS spectroscopic and imaging observations. A negative value means that the object was brighter at the epoch of the imaging observation than at the
epoch of the spectroscopic observation, and vice versa.}
\tablenotetext{c}{Elapsed time between the SDSS imaging and spectroscopic observations.}
\end{deluxetable}

\begin{deluxetable}{ccccccccc}
\tabletypesize{\scriptsize}
\tablecaption{Fe Coronal Line Fluxes\tablenotemark{a} \label{tab2}}
\tablewidth{0pt}
\tablehead{
\colhead{No} & \colhead{\fevii$\lambda$3759} & \colhead{\fevii$\lambda$5160} & \colhead{\fevii$\lambda$5722} &\colhead{\fevii$\lambda$6088} 
& \colhead{\fex} & \colhead{\fexi} & \colhead{\fexiv} & \colhead{$\log EM$\tablenotemark{b}}
}
\startdata
1& $<$18 & $<$12 & $<$21& $<$21 & 138$\pm$11& 111$\pm$21& 88$\pm$6 & 61.20 \\
2& 54$\pm$7& 30$\pm$10& 63$\pm$16& 97$\pm$9& 87$\pm$7& 55$\pm$8& 80$\pm$21 & 61.42 \\
3& 134$\pm$12& 41$\pm$17& 98$\pm$6& 129$\pm$6& 219$\pm$8& 164$\pm$9& 86$\pm$9 &61.70  \\
4& 89$\pm$8& 38$\pm$12& 59$\pm$5& 108$\pm$6& 69$\pm$5& 68$\pm$8& 37$\pm$10 & 61.22 \\
5& 47$\pm$5& $<$6& 35$\pm$4& 58$\pm$2& 94$\pm$4& 86$\pm$7& 28$\pm$3 & 61.06\\
6& $<$12& $<$18& $<$15& $<$9& 57$\pm$4& 72$\pm$5& 36$\pm$4 &60.66 \\
7& $<$9 & $<$9& $<$6& $<$6& 39$\pm$3& 50$\pm$5& 20$\pm$3 & 60.83\\
\enddata

\tablenotetext{a}{Emission line fluxes and 3$\sigma$ upper limits are in units of  $10^{-17}$ erg~cm$^{-2}$~s$^{-1}$, 
and only statistical errors are quoted.}
\tablenotetext{b}{Gas emissivity in units of cm$^{-3}$ obtained by summing over the emissivities 
of \fex, \fexi, and \fexiv\ divided by the abundance of Fe. A typical uncertainty is estimated 
to be from 0.1 to 0.3 dex without taking into account of ionization fraction corrections and 
uncertainties in the atomic data.}
\end{deluxetable}

\begin{deluxetable}{ccccccccccccc}
\tabletypesize{\scriptsize}
\tablecaption{Other Emission Lines$\tablenotemark{a}$} \label{tab3}
\tablewidth{0pt}
\tablehead{
\colhead{No} & \colhead{\hb$^n$} & \colhead{\hb$^b$} & \colhead{\ha$^n$} & 
\colhead{\ha$^b$} & \colhead{\heii} & \colhead{\nii} & \colhead{\sii} & 
\colhead{\siis} & \colhead{[O {\sc III}]$\lambda 5007$} &
\colhead{\neiii} 
}
\startdata
 1& 46$\pm$ 4& $<$33 &213$\pm$17& $<$57 & $<$9 & 54$\pm$ 3& 60$\pm$ 5& 25$\pm$2& 14$\pm$ 2&  $<$9\\
 2&384$\pm$ 7& $<$64 &1766$\pm$25& 100$\pm$33&12$\pm$ 5&695$\pm$20&247$\pm$ 5&171$\pm$4&452$\pm$23&119$\pm$15\\
 3&142$\pm$47&179$\pm$58&226$\pm$18&1592$\pm$36&89$\pm$13&110$\pm$16&101$\pm$10& 43$\pm$4&161$\pm$ 6& 54$\pm$11\\
 4& 53$\pm$19&499$\pm$37&336$\pm$51&1704$\pm$61&22$\pm$ 4& 74$\pm$ 7& 23$\pm$ 3& 26$\pm$3&219$\pm$ 6& 56$\pm$6\\
 5& 66$\pm$ 5& $<38$  &329$\pm$7& $<39$ &41$\pm$ 3& 40$\pm$ 5&  8$\pm$ 1& 10$\pm$2&212$\pm$ 4& 81$\pm$7\\
 6& 80$\pm$ 4& $<72$  &401$\pm$ 8& 67$\pm$22 &20$\pm$ 3& 78$\pm$ 4& 48$\pm$ 4& 32$\pm$3&104$\pm$ 3& $<$27\\
 7& 53$\pm$17& 83$\pm$24 &221$\pm$29&467$\pm$37& $<$9 & 96$\pm$ 9& 52$\pm$ 7& 24$\pm$3& 44$\pm$ 2&  $<$9
\enddata

\tablenotetext{a}{Emission line fluxes are in units of $10^{-17}$ erg~cm$^{-2}$~s$^{-1}$}
\end{deluxetable}

\begin{figure}
\plotone{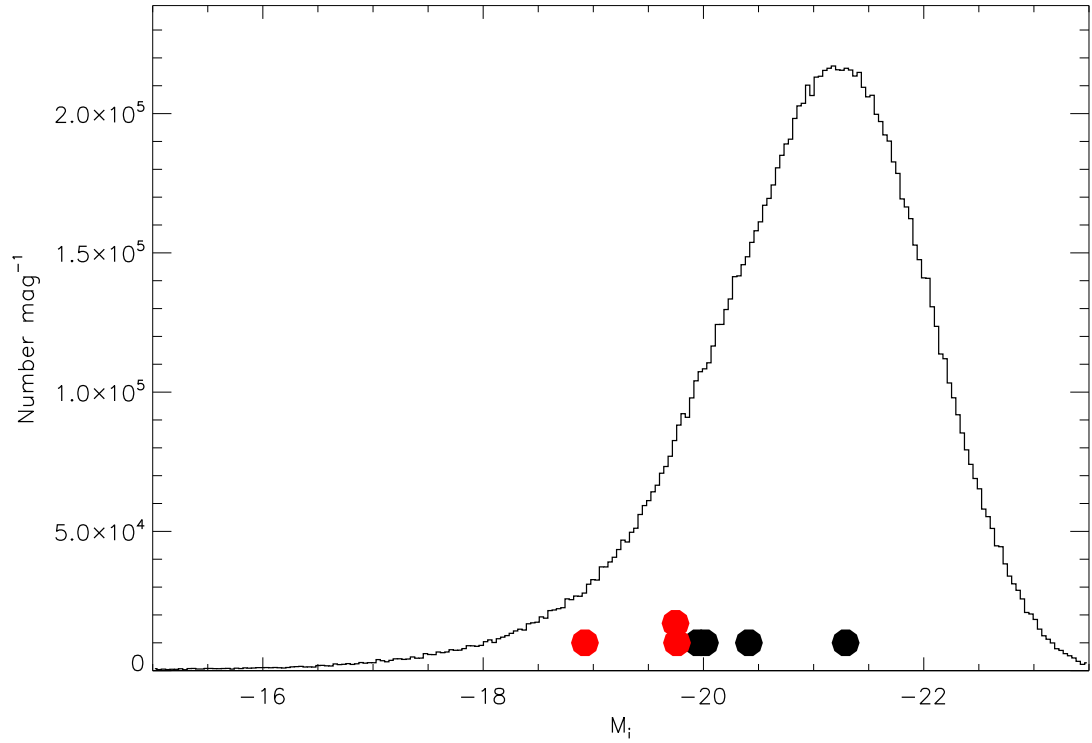}
\caption{Distribution of the $i$ band absolute magnitudes in the rest frame of the SDSS galaxies
that have spectral S/N ratio greater than 14.8, the lowest S/N ratio among the ECLE spectra in
this paper. The absolute magnitudes of the ECLEs are denoted as filled circles for those with \fevii\
 detected (red) and those without (black).}
\label{fig11}
\end{figure}

\begin{figure}
\epsscale{0.9}
\plotone{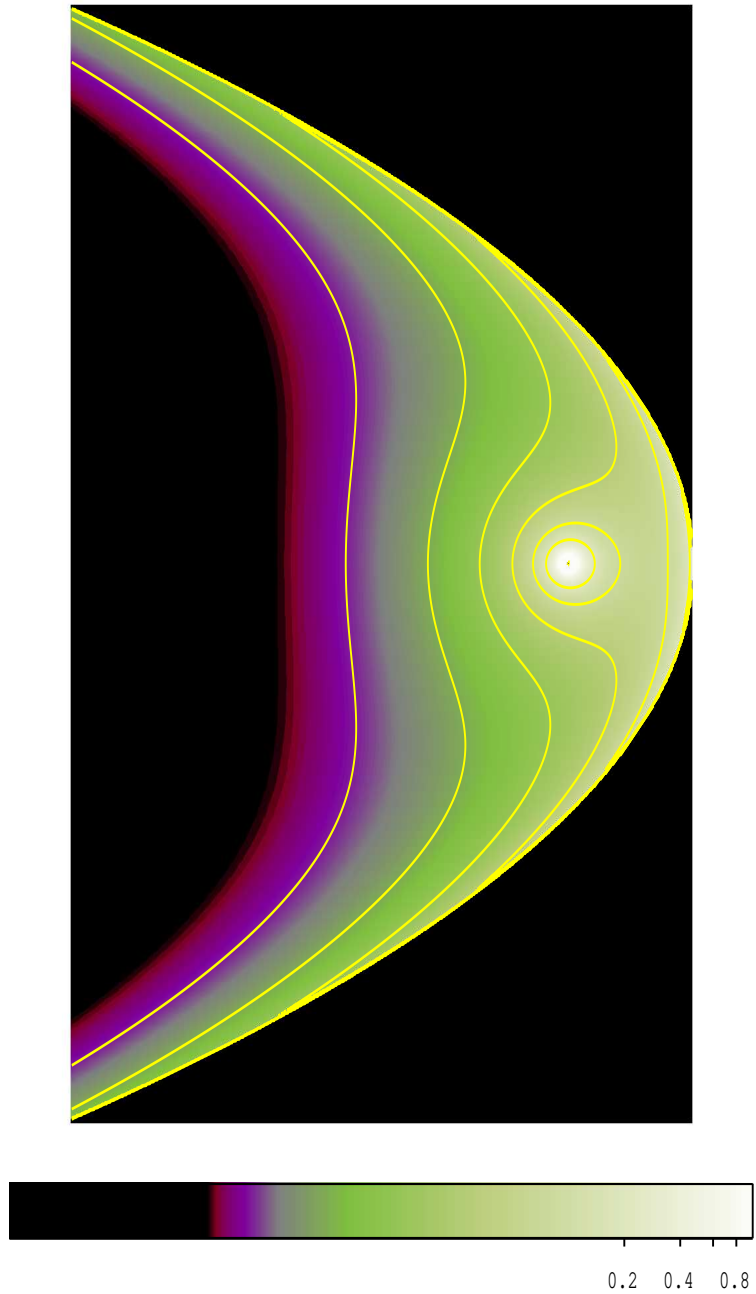}
\caption{Snapshot of the continuum intensity around a black hole that contributes to the 
observed emission lines at a specific time in an oversimplified model, in which the ionizing continuum
decreases with time as $\propto t^{-5/3}$ when $t > t_0$, where $t_0$ is the time of tidal disruption. The observer
is to the left of the figure at time $10 t_0$. The overlaid yellow lines are intensity contours with an
interval of 2.5 fold.}
\label{fig12}
\end{figure}       

\end{document}